\documentclass[jcp,reprint,a4paper,superscriptaddress]{revtex4-1}

\usepackage{amsmath,amssymb,amsthm,graphics}  

\theoremstyle{remark}

\usepackage[utf8]{inputenc}
\theoremstyle{definition}

\usepackage[pdftex,breaklinks,colorlinks]{hyperref}
\hypersetup{  
  pdftitle =
    {Norm conservation in Abedi, Maitra and Gross, PRL 105, 123002, 2010}
}
\hypersetup{  
    citecolor = blue,
    urlcolor = blue
}
\usepackage{graphicx}  
\usepackage{color}

\newcommand\argsR{\underline{\underline{\mathbf{R}}}}

\newcommand\argsr{\underline{\underline{\mathbf{r}}}}

\begin{document}

\allowdisplaybreaks[1]

\title{Comment on “Correlated electron-nuclear dynamics: Exact factorization of the molecular wavefunction” [J. Chem. Phys. 137, 22A530 (2012)]} 

\author{J. L. Alonso}
\affiliation{Departamento de F{\'{\i}}sica Te\'orica, Universidad de Zaragoza, 
Pedro Cerbuna 12, E-50009 Zaragoza, Spain}
\affiliation{Instituto de Biocomputaci\'on y F{\'{\i}}sica de Sistemas 
Complejos (BIFI), Universidad de Zaragoza, Mariano Esquillor s/n, Edificio 
I+D, E-50018 Zaragoza, Spain}
\affiliation{Unidad Asociada IQFR-BIFI, Mariano Esquillor s/n, Edificio 
I+D, E-50018 Zaragoza, Spain}
\author{J. Clemente-Gallardo}
\affiliation{Departamento de F{\'{\i}}sica Te\'orica, Universidad de Zaragoza, 
Pedro Cerbuna 12, E-50009 Zaragoza, Spain}
\affiliation{Instituto de Biocomputaci\'on y F{\'{\i}}sica de Sistemas 
Complejos  (BIFI), Universidad de Zaragoza, Mariano Esquillor s/n, Edificio 
I+D, E-50018 Zaragoza, Spain}
\affiliation{Unidad Asociada IQFR-BIFI, Mariano Esquillor s/n, Edificio 
I+D, E-50018 Zaragoza, Spain}
\author{P. Echenique-Robba}
\affiliation{Instituto de Qu\'{\i}mica F\'{\i}sica Rocasolano, CSIC, Serrano 
119, E-28006 Madrid, Spain}
\affiliation{Departamento de F{\'{\i}}sica Te\'orica, Universidad de Zaragoza, 
Pedro Cerbuna 12, E-50009 Zaragoza, Spain}
\affiliation{Instituto de Biocomputaci\'on y F{\'{\i}}sica de Sistemas 
Complejos (BIFI), Universidad de Zaragoza, Mariano Esquillor s/n, Edificio
I+D, E-50018 Zaragoza, Spain}
\affiliation{Zaragoza Scientific Center for Advanced Modeling (ZCAM), 
Universidad de Zaragoza, Mariano Esquillor s/n, Edificio I+D, E-50018 
Zaragoza, Spain}
\affiliation{Unidad Asociada IQFR-BIFI, Mariano Esquillor s/n, Edificio 
I+D, E-50018 Zaragoza, Spain}
\author{J. A. Jover-Galtier}
\affiliation{Departamento de F{\'{\i}}sica Te\'orica, Universidad de Zaragoza, 
Pedro Cerbuna 12, E-50009 Zaragoza, Spain}
\affiliation{Instituto de Biocomputaci\'on y F{\'{\i}}sica de Sistemas 
Complejos (BIFI), Universidad de Zaragoza, Mariano Esquillor s/n, Edificio 
I+D, E-50018 Zaragoza, Spain} 
\date{\today}

\begin{abstract}

In spite of the relevance of the proposal introduced in the recent work
A.~Abedi, N.~T.~Maitra and E.~K.~U.~Gross, \emph{J. Chem. Phys.} \textbf{137},
22A530, 2012, there is an important ingredient which is missing. Namely, the
proof that the norms of the electronic and nuclear wavefunctions which are the
solutions to the nonlinear equations of motion are preserved by the evolution.
To prove the conservation of these norms is precisely the objective of this
Comment.

\end{abstract}
\maketitle


In a remarkable recent work \cite{Abedi2012}, Abedi et al. present an exact
factorization of the molecular wavefunction into a nuclear and an electronic
part, which allows to rigorously introduce generalized and very useful
concepts, such as the time-dependent potential energy surface. This formalism
also sets the stage to better understand, and hence probably improve, very
much used quantum-classical schemes, such as Ehrenfest, surface-hopping or
Born-Oppenheimer dynamics.

In spite of the relevance of the proposal, we consider that there is an
important ingredient which is missing. Namely, the proof that the norms of the
two functions $\Phi_{\argsR}(\argsr, t)$ and $\chi(\argsR, t)$, which are
solutions of the nonlinear Eqs.~(28) and~(29) in \cite{Abedi2012}, are
conserved. This is a key point in order to associate $\Phi_{\argsR}(\argsr,
t)$ and $\chi(\argsR, t)$ to a marginal and a conditional probability
amplitude, respectively, thus leading to their identification as nuclear and
electronic wavefunctions, as it is the purpose of \cite{Abedi2012}.

In \citep{Abedi2012}, it is proved that, given an exact solution,
$\Psi(\argsR, \argsr,t)$, of the time-dependent molecular Schr\"odinger
equation, it can be written as a single product of the form
\begin{equation}
  \label{eq:1}
  \Psi(\argsR, \argsr,t)=\Phi_{\argsR}(\argsr, t)\chi(\argsR,t) \ , 
\end{equation}
such that the \textit{partial normalization condition} (PNC),
\begin{equation}
  \label{eq:3}
  \int d\argsr|\Phi_{\argsR}(\argsr, t)|^2=1 \qquad \forall \, \argsR, t \ ,
\end{equation}
is satisfied. This condition implies that also $\chi$ is normalized if $\Psi$
is. This is proved by providing a constructive definition of
$\Phi_{\argsR}(\argsr, t)$ and $\chi(\argsR, t)$ in terms of $\Psi(\argsR,
\argsr,t)$ in Eqs.~(25) and (26). However, when the equations of motion for
the former are variationally obtained, their constructive definition is not
used, thus requiring an independent proof that the PNC holds for the 
solutions.

If one wants to be sure that a given set of equations of motion do conserve
some quantity, there are essentially two options: Either one explicitly forces
the conservation at the action level, e.g., using Lagrange multipliers, or one
shows that there is another reason (e.g., a symmetry of the action) why the
equations of motion produce the conservation. In \cite{Abedi2012}, neither of
these two things are explicitly done. Instead, the PNC is only used to
simplify the Euler-Lagrange equations once they have been obtained from the
stationary action principle. This occasional use of the PNC does not
guarantee, in principle, that it holds for all times if no further proof is
provided.

But, before detailing our proof of this property does hold, let us point out that
two possible ways of proving it have been discarded here for different
reasons. First, one could have shown that the action has a certain symmetry
and obtain the conservation law as an application of Noether's theorem. We
have been unable to find such a symmetry. Second, notice that the equations of
motion in \cite{Abedi2012} can be written as
\begin{subequations}
\label{eq:EoM}
\begin{align}
i \partial_{t}\Phi_{\argsR}(\argsr, t) & = 
 \hat{H}_\Phi[\Phi_{\argsR},\chi,\partial_{t}\Phi_{\argsR}] \,
 \Phi_{\argsR}(\argsr, t) \ , \label{eq:EoM_Phi} \\
i \partial_{t}\chi(\argsR, t) & = 
 \hat{H}_\chi[\Phi_{\argsR},\chi,\partial_{t}\Phi_{\argsR}] \,
 \chi(\argsR, t) \ . \label{eq:EoM_chi}
\end{align}
\end{subequations}
If the operators $\hat{H}_\Phi$ and $\hat{H}_\chi$ were linear and Hermitian,
the conservation of the norm of the functions $\chi$ and $\Phi_{\argsR}$ would
be straightforward. As that is not the case, a more careful analysis is in
order, but a formal proof using this idea is sometimes useful if one assumes
that all necessary conditions on the corresponding infinite dimensional
Hilbert spaces are satisfied. However, in this particular case, and due to the
dependence of the time derivative of $\Phi_{\argsR}(\argsr, t)$ in the
definition of the operators, the only way to have Hermiticity is to show in
advance that the norm of $\Phi_{\argsR}(\argsr, t)$ is indeed conserved. This
renders the reasoning circular, and thus invalid.

The problem of obtaining the dynamical equations can be addressed from a
different perspective. Consider then a decomposition of the form given by
Eq.~(\ref{eq:1}) but without imposing the PNC condition. In this case, we
define:
\begin{equation}
  \label{eq:4}
  f(\argsR, t) := \int d\argsr |\Phi_{\argsR}(\argsr,t)|^{2} \ ,
\end{equation}
and, if we use that $\Psi$ is normalized at all times, we have:
\begin{equation}
  \label{eq:5}
  \int d\argsR f(\argsR, t)|\chi(\argsR,t)|^{2}= 1 \ .
\end{equation}

If we consider this factorization for $\Psi$ and introduce it in the
variational framework used in \cite{Abedi2012}, we obtain as dynamical
equations: 
\begin{align}
\label{eq:6}
  & if \partial_{t}\chi= \Big[ f (\hat T_{n} + \hat V_{n}) 
  -i \sum_{\alpha=1}^{N_{n}}\frac 1 M_{\alpha}(\vec {\mathbf{A}}_{\alpha}\cdot
    \vec \nabla_{\alpha}) \nonumber \\
 & \quad + \langle \Phi_{\argsR}|\hat T_{e} + \hat T_{n} + \hat V_{e} + \hat W_{en} - 
     i\partial_{t}|\Phi_{\argsR}\rangle \Big] \chi \ , \\
  \label{eq:7}
  & if \partial_{t}\Phi_{\argsR}|\chi|^{2} =  
\Bigg[|\chi|^{2} \Big(  f [ \hat T_{e} +
    \hat T_{n} +\hat V_{e} + \hat W_{en} ] \nonumber \\
 & \quad - \langle \Phi_{\argsR} \large | \big( \hat T_{e} +
   \hat  T_{n}  +\hat V_{e} + \hat W_{en} - i\partial_{t} \big)
   |\Phi_{\argsR}\rangle  \Big) - \nonumber \\
& \quad \chi^{*} \sum_{\alpha=1}^{N_{n}}\frac
  1{M_{\alpha}} \Big( f(\vec \nabla_{\alpha}\chi)\vec
      \nabla_{\alpha} + i \vec A_{\alpha}\cdot
      (\vec \nabla_{\alpha}\chi) \Big) \Bigg]
\Phi_{\argsR} \ ,
\end{align}
where the dependencies have been omitted, and
\begin{align*}
  \hat V_{e}(\argsr,t):&=\hat W_{ee}(\argsr)+\hat V_{ext}^e(\argsr, t) \ ,
\\
  \hat V_{n}(\argsR,t):&=\hat W_{nn}(\argsR)+\hat V_{ext}^n(\argsR, t) \ .
\end{align*}

Any factorization of the form~(\ref{eq:1}), with $\Psi$ a solution of the
molecular Schrödinger equation, satisfies~(\ref{eq:6}) and ~(\ref{eq:7}).
Also, notice that, if we have $f(\argsR, t)=1$ for all $t$ (i.e., the
factorization satisfies the PNC for all values of time), then these equations
reduce to Eqs.~(28) and~(29) in \cite{Abedi2012}. Now, the factorization
in~(\ref{eq:1}) exhibits an invariance under the group of invertible functions
on $\mathbb{C}_{0}=\mathbb{C}-\{ 0\}$:
\begin{equation}
\label{eq:7b}
\tilde \chi (\argsR, t)=a(\argsR, t) \chi(\argsR, t) \ , \
\tilde \Phi_{\argsR}(\argsr, t)=\Phi_{\argsR}(\argsr, t)/ a(\argsR, t) ,
\end{equation}
where $a(\argsR, t)$ is any complex function without zeros. In other words,
for any given solution, $\chi (\argsR, t)$ and $\Phi_{\argsR}(\argsr, t)$,
of~(\ref{eq:6}) and~(\ref{eq:7}), we can obtain new solutions, $\tilde \chi
(\argsR, t)$ and $\tilde \Phi_{\argsR}(\argsr, t)$, which produce the same
$\Psi(\argsR, \argsr, t)$, by applying the above transformation. Of course,
these new functions will be solution to the equations with the corresponding
$\tilde f (\argsR, t)$. Also notice that this `gauge freedom' enlarges the
$U(1)$--freedom discussed in \cite{Abedi2012}, where only the phase of each
function is transformed.

Let us consider now a gauge fixing defined as
\begin{equation}
  \label{eq:2}
  a(\argsR, t)=e^{i\theta(\argsR,t)}\sqrt{f(\argsR, t)}
\end{equation}
where $f(\argsR, t)$ is defined by Equation (\ref{eq:4}) and $\theta (\argsR,
t)$ is arbitrary. The transformation is considered for the full trajectory
$\Psi(\argsR, \argsr, t)$, since it depends explicitly on the norm of the
function $\Phi_{\argsR}$ along it. In particular, if we consider a solution
with initial unit norm, i.e., $f (\argsR, 0) = 1$, $\forall \argsR$, we find
that the initial conditions for the original and the transformed curves
coincide:
\begin{equation}
  \label{eq:8}
  \tilde \chi(\argsR, 0)=\chi(\argsR, 0)\ , \qquad 
\tilde \Phi_{\argsR}(\argsr, 0)=\Phi_{\argsR}(\argsr, 0) \ ,
\end{equation}
and that functions $\tilde \Phi_{\argsR}$  and $\tilde \chi$ are
normalized by construction for all values of time:
\begin{equation}
  \label{eq:9}
  \int d\,\argsR |\tilde \chi(\argsR, t)|^{2}=1=\int d\argsr |\tilde
  \Phi_{\argsR}(\argsr, t)|^{2} \qquad \forall t, \argsR \ .
\end{equation}

Finally, consider any solution of Eqs.~(28) and~(29) in \cite{Abedi2012} for
some initial conditions $\chi^{0}(\argsR)$ and $\Phi^{0}_{\argsR}(\argsr)$
that satisfy the PNC, and let us ask whether or not the PNC is satisfied at
subsequent times. We have seen that, among the factorizations of the molecular
wavefunction $\Psi(\argsR, \argsr, t)$ [with initial conditions $\Psi(\argsR,
\argsr, 0)=\chi^{0}(\argsR)\Phi_{\argsR}^{0}(\argsr)$)] there is one, given by $\tilde \chi(\argsR,t)$ and $\tilde \Phi_{\argsR}(\argsr, t)$,
which satisfies the PNC for all values of time, and also Eqs.~(\ref{eq:6}--\ref{eq:7}) with $\tilde f(\argsR, t)=1$ and
initial conditions $\chi^{0}(\argsR)$ and $\Phi^{0}_{\argsR}(\argsr)$. Now, given that Eqs.~(\ref{eq:6}--\ref{eq:7}) with
$\tilde f(\argsR, t)=1$ are precisely Eqs.~(28) and~(29) in \cite{Abedi2012},
we have that the functions $\tilde \chi(\argsR, t)$ and $\tilde
\Phi_{\argsR}(\argsr, t)$ are a also a solution to them, with initial
conditions $\chi^{0}(\argsR)$ and $\Phi_{\argsR}^{0}(\argsr)$, and such that
their norms are preserved for all time. Since we have not modified the initial
conditions, if we assume that Eqs.~(28) and~(29) in \cite{Abedi2012} have
unique solution for given initial conditions, we can conclude that the
norm-conserving solution to them that we have found must be the arbitrary one
we began with, thus showing that every
solution of Eqs.~(28) and~(29) in \cite{Abedi2012} with initial conditions
that satisfy the PNC also satisfies it at all times.

\vspace{8pt}
After submitting this comment for publication, one of the authors of
\cite{Abedi2012} as well as the referee that reviewed our work brought to our
attention the alternative (unpublished) proof based on the
method of Lagrange multipliers.

\section*{Acknowledgements}

We would like to thank Dr. Alberto Castro for many useful discussions. 
This work has been supported by grants FIS2009-13364-C02-01
(MICINN, Spain), and Grants E24/1 and E24/3 (DGA, Spain).

\bibliographystyle{apsrev4-1}


\begin{thebibliography}{2}%
\makeatletter
\providecommand \@ifxundefined [1]{%
 \@ifx{#1\undefined}
}%
\providecommand \@ifnum [1]{%
 \ifnum #1\expandafter \@firstoftwo
 \else \expandafter \@secondoftwo
 \fi
}%
\providecommand \@ifx [1]{%
 \ifx #1\expandafter \@firstoftwo
 \else \expandafter \@secondoftwo
 \fi
}%
\providecommand \natexlab [1]{#1}%
\providecommand \enquote  [1]{``#1''}%
\providecommand \bibnamefont  [1]{#1}%
\providecommand \bibfnamefont [1]{#1}%
\providecommand \citenamefont [1]{#1}%
\providecommand \href@noop [0]{\@secondoftwo}%
\providecommand \href [0]{\begingroup \@sanitize@url \@href}%
\providecommand \@href[1]{\@@startlink{#1}\@@href}%
\providecommand \@@href[1]{\endgroup#1\@@endlink}%
\providecommand \@sanitize@url [0]{\catcode `\\12\catcode `\$12\catcode
  `\&12\catcode `\#12\catcode `\^12\catcode `\_12\catcode `\%12\relax}%
\providecommand \@@startlink[1]{}%
\providecommand \@@endlink[0]{}%
\providecommand \url  [0]{\begingroup\@sanitize@url \@url }%
\providecommand \@url [1]{\endgroup\@href {#1}{\urlprefix }}%
\providecommand \urlprefix  [0]{URL }%
\providecommand \Eprint [0]{\href }%
\providecommand \doibase [0]{http://dx.doi.org/}%
\providecommand \selectlanguage [0]{\@gobble}%
\providecommand \bibinfo  [0]{\@secondoftwo}%
\providecommand \bibfield  [0]{\@secondoftwo}%
\providecommand \translation [1]{[#1]}%
\providecommand \BibitemOpen [0]{}%
\providecommand \bibitemStop [0]{}%
\providecommand \bibitemNoStop [0]{.\EOS\space}%
\providecommand \EOS [0]{\spacefactor3000\relax}%
\providecommand \BibitemShut  [1]{\csname bibitem#1\endcsname}%
\let\auto@bib@innerbib\@empty
\bibitem [{\citenamefont {Abedi}\ \emph {et~al.}(2012)\citenamefont {Abedi},
  \citenamefont {Maitra},\ and\ \citenamefont {Gross}}]{Abedi2012}%
  \BibitemOpen
  \bibfield  {author} {\bibinfo {author} {\bibfnamefont {A.}~\bibnamefont
  {Abedi}}, \bibinfo {author} {\bibfnamefont {N.~T.}\ \bibnamefont {Maitra}}, \
  and\ \bibinfo {author} {\bibfnamefont {E.~K.~U.}\ \bibnamefont {Gross}},\
  }\href {\doibase 10.1063/1.4745836} {\bibfield  {journal} {\bibinfo
  {journal} {J. Chem. Phys.}\ }\textbf {\bibinfo {volume} {137}},\ \bibinfo
  {pages} {22A530} (\bibinfo {year} {2012})}\BibitemShut {NoStop}%
\end{thebibliography}


\end{document}